\documentclass[a4paper,10pt]{article}

\usepackage{cmap}
\usepackage[T2A]{fontenc}
\usepackage[utf8]{inputenc}
\usepackage[english]{babel}
\usepackage{amsmath}
\usepackage{amssymb}
\usepackage{xcolor}
\usepackage{braket}
\usepackage{graphicx}
\usepackage{bm}
\usepackage{tikz}
\usepackage{authblk}

\usepackage[left=2cm,right=2cm,top=3cm,bottom=3cm]{geometry}
\tikzset{every picture/.style={line width=0.75pt}} 

\author[1,2]{K.V.~Nikolaev}
\author[3]{V.~Soltwisch}
\author[4]{M.A.~Botchev}
\author[3]{A.~Fernández Herrero}
\author[3]{P.~H{\"o}nicke}
\author[3]{F.~Scholze}
\author[1]{S.N.~Yakunin} 
\affil[1]{National Research Center Kurchatov Institute, Moscow, Russia}
\affil[2]{Moscow Institute of Physics and Technology, Dolgoprudny, Russia}
\affil[3]{Physikalisch-Technische Bundesanstalt (PTB), Berlin, Germany}
\affil[4]{Keldysh Institute of Applied Mathematics of RAS, Moscow, Russia}

\title{\bf Polygon-based unified Fourier-modal approach \\ for diffractive optics simulations}

\date{July 1, 2024}

\begin{document}

\maketitle
\begin{abstract}
In this article, we derive a theoretical formalism that unifies the rigorous coupled wave analysis and the dynamical diffraction theory. Based on this formalism, we design a computational approach for the diffraction calculation for the nanoscale lamellar gratings with an arbitrary line profile shape. In this approach, the gratings line profile is approximated as a polygon. This proves to be convenient since such an approach does not rely on the geometry model of the grating. We test the new approach against other computational theories and a synchrotron scattering experiment.
\end{abstract}

\section{Introduction}

The problem of monochromatic wave scattering on a periodic potential arises in many fields of science and technology.
Starting with early X-ray diffraction on crystals research,
it is now in photonics~\cite{ikonnikov2023fork} and plasmonics~\cite{sharma2020Au},
engineering of astronomical equpment~\cite{muslimov2023optical},
and diffractive optics,
which, for instance, now have applications in augmented~reality devices~\cite{lim2021augmented}.
The synchrotron nano-metrology is
yet another research field in which the scattering on periodic systems is employed.
In the context of synchrotron nano-metrology,
EUV and X-ray scattering is used to characterize the structure of nano-systems.
For instance, the pattern recorded in
the grazing-incidence small-angle scattering experiments
can be used to reconstruct the shape of lamellar gratings~\cite{soltwisch2017reconstructing}
with precision on the sub-nanometer scale.
It was also experimentally shown
that the diffuse scattering pattern encodes the statistical parameters of the gratings wall roughness
(line edge/width roughness)~\cite{fernandez_herrero2017charachteristic, fernandez_herrero2022analysis}.
Thus, scattering can be used not only to characterize the averaged parameters of the nanostructure
but also to inspect the statistics of its surface imperfections.
EUV scatterometry has its main applications in microelectronics as the mirrors used in EUV lithography today are based on Mo/Si multilayer systems optimized for maximum reflectivity at 13.5 nm.
Such research is currently of great interest,
and we see three distinct reasons for this.
First, the commercial success of the EUV lithography.
Second, recent developments in high-harmonic EUV generation~\cite{ku2016scatterometer,dombi2023}
may allow the EUV scatterometry in-lab, which until now has mainly been possible on synchrotrons.
Finally, the third reason is that modern elements of microelectronics,
like gate-all-around transistors,
have a complex architecture that necessitates an advanced 3D metrology.
Such sophisticated manufacturing technology requires modeling of the scattering on the complex shape structures,
which in turn motivates improvement of the computational schemes.
This describes just one use case in which solving the wave scattering problem is essential,
and there are many more.

We focus on solving the wave equation for periodic scattering potential.
Several theoretical and numerical approaches exist for solving this type of problem.
Among them are the finite element method~\cite{pomplun2007adaptive},
the finite difference time domain approach~\cite{taflove2013advances},
methods based on boundary integral equations~\cite{goray2010solving}
or surface coordinate transform (C-method)~\cite{chandezon1980new},
and the Fourier modal methods.
Fourier modal methods exploit the periodicity of a structure's scattering potential to formulate the solution in terms of the sum of the plane waves.
As mentioned, the problem of scattering on a periodic potential arises in different fields of science.
Thereby, there are several instances of Fourier modal theories,
which historically developed in parallel with little to no overlap in the literature.
The scientific community around diffractive optics and holography uses the rigorous coupled wave analysis (RCWA) theory~\cite{kogelnik1969coupled}.
The synchrotron radiation community employs the dynamic diffraction theory
(DDT)~\cite{kaganer1982transition, batterman1964dynamical}.
Essentially, these theories are identical.
The central idea is to represent the scattering potential as the discrete Fourier series and then to search for the solution in terms of the Bloch series.
This allows the reduction of the wave equation, which is inconveniently a partial differential equation,
to a system of ordinary differential equations, which one can solve in terms of the eigenvalue problem.
This idea constitutes both the RCWA and the DDT.
However, there is a difference in how these theories are developed,
which is due to the specifics of each of the application fields.
For example, constructing a Bloch series of only two components,
a reflected wave and a diffraction order of interest, is often sufficient in the X-ray range.
Indeed, the wavelength for X-rays is shorter than in the optical range, and therefore,
a phase difference between different plane waves is higher.
Hence, interference between the diffraction orders is a rare occurrence for X-rays.
In contrast, in the optical range one typically has to consider the multiple wave interference to solve the wave equation with satisfying accuracy.
That is why RCWA is formulated to account for an arbitrary number of wave modes, whereas in DDT,
one typically finds an explicit solution for each wave mode.
On the other hand, the explicit form of DDT allows to account for any orientation of periodic media,
while RCWA assumes only a laterally periodic structure.
Again, this difference is driven by applications:
diffraction gratings that are periodic along the surface of the structure for RCWA,
and that is crystals periodic in three dimensions for DDT.
In RCWA, the complex line profile of gratings is approximated with a series of layers,
homogeneous in the vertical direction.
Then, the solution in each layer is chosen so that the particular solution across the whole structure is continuous.

In summary, RCWA and DDT are similar theories.
The RCWA considers many wave interference but works only on vertically homogeneous structures,
while DDT considers structures with an arbitrarily oriented periodicity.
The similarity between these approaches was mentioned in early works~\cite{kogelnik1969coupled}.
Therefore, the idea of combining the two approaches,
so that both multiple wave interference and complex structure are within reach,
is a natural one.
One way to implement this was first demonstrated in~\cite{colella1974multiple} and further developed in~\cite{stetsko1997algorithm}.
In considerably recent research~\cite{huang2013fourier},
the idea of unifying both theories was revisited.
In addition,
a new idea for treating the diffraction gratings with complex profiles
without slicing them on vertically homogeneous layers was proposed~\cite{huang2013fourier}.

With this article,
we aim to once again draw the attention of the optical and the synchrotron scientific communities
to the fundamental similarity of RCWA and DDT.
Based on the ideas
in~\cite{colella1974multiple,
stetsko1997algorithm,
huang2013fourier,
stepanov1998dynamical,
mikulik1999xray},
we construct yet another unified Fourier modal theory.
We do this in a simplistic manner,
which allows us to focus on the main ideas essential for accounting for multiple wave interference in structures with complex profiles.
For this theory,
we derive the generalized boundary conditions in linear algebraic terms
so that it would be possible to implement them numerically in a computationally efficient way in the future.
Finally, we implement a polygonal approach to approximate the shape of grating profiles.
This opens a direction for future research towards model-independent simulations of diffraction gratings.
We test our approach against the conventional RCWA and the finite element method.
We also test the approach to analyze actual experimental data previously published
in~\cite{soltwisch2017reconstructing}.
The numerical and experimental models presented in this article are relevant for the further development of EUV or soft X-ray nanometrology.
However, it should be emphasized that the actual applications go beyond this framework.

\section{Theory}

First, let us describe the mathematical notation used in this article. 
Fourier modal methods describe waves in terms of amplitude vectors,
which are vectors of Hilbert space.
Amplitude vectors and matrices are written in capital letters in normal font, for example $A$.
The space vectors, which describe the real geometric objects,
are written in lowercase bold letters,
for example $\bm{r}$,
to avoid confusion with amplitude vectors.
The only exception to the capital letter notation is the electric field.
The notation for the electric field is $E(\bm r)$,
which is well established in the literature.
The Fourier components of $E(\bm r)$ are written as $\hat E_g$ and they constitute the amplitude vector $E$.
The electric field in real space $E(\bm r)$ is always explicitly written as a function of coordinate
to avoid confusing it with the amplitude vector $E$.
In this article we neglect polarization effects, so $E(\bm r)$ is a scalar field.

We start with the Helmholtz equation in the scalar field approximation:
\begin{equation}
	(\Delta + k^2) E (\bm r) = -k^2\chi(\bm r) E(\bm r)
        .
	\label{eq:helmholtz}
\end{equation}
The scalar approximation is usually valid in the X-ray and EUV region for sufficiently small grazing incidence angles.
The validity of the approximation depends on the structure of the scattering potential,
see validity criteria in~\cite{goray1994nonscalar, goray2005scalar}.
Here, $k$ is the wave number in the vacuum and $\chi$ is the dielectric susceptibility,
the term $-k^2\chi$ represents the scattering potential in case of the photons.
The dielectric susceptibility function is linearly dependent on the electron density distribution,
thus it encodes the geometry of the nanostructure.
Since we consider periodic nanostructures, $\chi(\bm r)$ can be represented as a Fourier series:
\begin{equation}
	\chi(\bm r) = \sum_g \chi_g e^{i\bm{g} \bm{r} }
	.
	\label{eq:chi}
\end{equation}
For brevity,
we allow ourselves an ambiguous use of the index notation:
$\bm g$ is the reciprocal space vector.
The subscript $g$ in the sum in Eq.~\ref{eq:chi}
implies the sum over all nodes in the reciprocal space.
In general, we will consider a two-dimensional periodic structure $\bm g = [2\pi m/d_x,\; 2\pi n/d_z]^T$,
where $m,n \in \mathbb{Z}$ and $d_{x,z}$ are the elementary cell sizes.

The solution of a wave equation with a periodic scattering potential can be expressed as a series of plane waves by the Bloch theorem.
Thus, an ansatz
for Eq.~\ref{eq:helmholtz} is
\begin{equation}
	E(\bm r) =  \sum_g \hat E_g e^{i\bm k_g \bm r}
	.
	\label{eq:bloch}
\end{equation}
In the far field
(in the ambient at a sufficient distance from the surface of the sample), 
the phases of these plane waves are related with the nodes in reciprocal space $\{\bm g\}$ by the Laue diffraction condition
$\bm k_g = \bm k + \bm g$, where $\bm k$ is the wave-vector of the incident rays in vacuum.
Thus, $\bm k_g$ is the wave vector for the $g$-th wave mode (or $g$-th diffraction order).
Direct substitution of Eqs.~\ref{eq:bloch}~ and~\ref{eq:chi} in the Eq.~\ref{eq:helmholtz}
yields a system of linear equations:
\begin{equation}
	\left( k_g^2 - k^2\right) \hat E_g
        =
        k^2 \sum_h \chi_{g-h} \hat E_h
	.
	\label{eq:disper}
\end{equation}
This is an infinite system of equations, for each node denoted by $g$ each such equation contains infinite Fourier components of $\chi$.
The system is derived with use of the phase shift property of the Fourier series:
$\sum_h \chi_{g-h} = \sum_h \chi_g e^{i\bm h \bm r} $,
for any arbitrary constant $\bm r$.

The physical meaning of Eq.~\ref{eq:disper} becomes apparent if we rearrange it.
In the sum on the right-hand side of the Eq.~\ref{eq:disper},
consider the term $\chi_0\hat E_g$ that corresponds to $g=h$.
Let us move it to the left hand-side.
Then, by introducing $\kappa^2 = (1+\chi_0) k^2$ one can rewrite Eq.~\ref{eq:disper} as follows:
\begin{equation}
	\left( k_g^2 - \kappa^2\right) \hat E_g
        =
        k^2 \sum_{h \neq g} \chi_{g-h} \hat E_h
	,
        \label{eq:disper_2}
\end{equation}
The Fourier component $\chi_0$ represents the average dielectric permittivity of the structured medium.
Therefore, $\kappa$ can be interpreted as the wave number within the structure 
as if it were an unstructured continuous medium
(effective homogeneous medium).
If the wavenumber $k_g$ of a diffracted wave $k_g = \kappa$,
then for higher orders $g \neq 0$ and
Eq.~\ref{eq:disper_2} has only trivial solutions $\hat E_g = 0$.
This has two consequences. 
First is a trivial one:
if the medium is indeed homogeneous,
then $k_g = \kappa$ and there is only a specularly reflected wave $\hat E_0$.
Second, in the structured medium,
diffraction is only possible if $k_g$ deviates from $\kappa$.
Thus, Eq.~\ref{eq:disper} describes dispersion in the structured medium.

To calculate the dispersion, the symmetry of the scattering potential can be taken into account: all wavevectors should be invariant with respect to translations along the surface~\cite{landau}:
$\bm{k}_{g\parallel} = \bm{k}_{0 \parallel} + \bm{g}_\parallel \equiv \text{const}$.
Then, the aberration of $k_g$ is only due to a change in its vertical component $k_{gz}$~\cite{afanasev1983x} for which we can write
$(k_g^2 - k^2) = (k_{gz} + g_z)^2 - \kappa_{gz}^2 $,
where 
%
\begin{equation}
	\kappa_{gz}^2 = (1+\chi_0)k^2
        -
        (\bm{k}_{\parallel} + \bm{g}_{\parallel})^2 
	,
	\label{eq:spherical}
\end{equation}
and the $k_{gz}^2$ is a value to be sought.
Eq.~\ref{eq:spherical} is the equation of a sphere and it describes the dispersion in the effective medium (spherical dispersion equation).
In the trivial case of homogeneous media, it results in the Snell's law.
Now, introducing normalization $\xi = k_{gz}/k$, Eq.~\ref{eq:disper} take a form
\begin{equation}
        \left(
        \xi^2 + \dfrac{2 g_z}{k} \xi
        + \dfrac{g^2_z-\kappa^2_{gz}}{k^2}
        \right)
        \hat E_g
        =
        \sum_h \chi_{g-h} \hat E_h
	.
	\label{eq:disper_explicit}
\end{equation}
This is an infinite system of quadratic equations with respect to $\xi$ with infinite terms on the right-hand side.
We note once again that $\xi$ corresponds to the vertical components of the wave vectors. 
Note that the index $g$ runs over an equation in a system and the index $h$ corresponds to the term in the sum.
We can give a physical interpretation to this:
the index $g$ corresponds to a wave mode $\hat E_g$, and $h$ enumerates a vector from the basis (an eigenstate vector), a linear combination of which forms $\hat E_g$ itself.
In these linear combinations,
the components $\chi_{g-h}$ play the role of coupling coefficients.
Since $\chi(\bm{r})$ is $\mathbb{L}^2$ integrable,
the larger the separation $|g-h|$,
the smaller the coupling.
It is more convenient to analyze such systems using linear algebra.

In the language of linear algebra, Eq.~\ref{eq:disper_explicit} have a form
\begin{equation}
	\left[ \xi^2 I + \xi A + B \right] E = 0
	.
	\label{eq:qep}
\end{equation}
Here, $I$ is the identity matrix,
$A = \text{diag}(\ldots 2g_z/k \ldots) $
and
$B = \text{diag}(\ldots [g_z^2 - \kappa_{gz}^2]/k^2 \ldots) - C_{[\chi]}$.
Last term here is a Toeplitz matrix~\cite{gray2006toeplitz}
filled with the Fourier components $\chi_{g}$:
\begin{equation}
	C_{[\chi]} =
	\left[
		\begin{array}{lllll}
			\ddots & & & &  \\
			& \chi_0 & \chi_{\overline 1} & \chi_{\overline 2} &  \\
			& \chi_1 & \chi_0 &    \chi_{\overline 1} &  \\
			& \chi_2 & \chi_{1} &  \chi_{0} &  \\
			& & & & \ddots \\
		\end{array}
	\right]
	.
	\label{eq:toeplitz}
\end{equation}
Note again that we are considering a 2D structure,
and each component $\chi_g$ corresponds to a node $g$ in the reciprocal space
which inherently has two coordinates,
but for the brevity of notation we use the single index $g$.
To construct the matrix $C_{[\chi]}$,
a 2D grid of nodes $\{\bm g\}$ is unfolded in a 1D array in any arbitrary order,
it is only important to maintain same ordering for the other matrices involved in computations.
A Toeplitz matrix $C_{[\chi]}$ encodes information about the scattering potential:
the more Fourier components,
i.e. more nodes in reciprocal space are taken into account,
the more detailed the structure is obtained. 
Let $n$ be the number of nodes, so that
$C_{[\chi]} \in \mathbb{C}^{n \times n}$.
The matrix $A$ contains vertical coordinates $g_z$
of the nodes in the reciprocal space and the matrix $B$ encodes the  scattering potential structure
and its diagonal part takes into account the spherical dispersion.
The Eq.~\ref{eq:qep} represents dispersion inside the structured medium.
It has the form of a quadratic eigenvalue problem (QEP) with respect to the eigenvalue $\xi$ and eigenvector $E$.

Consider the case of a vertically homogeneous structure:
for instance a box-shaped grating.
The structure is constant with respect to the vertical translation, so there are no reciprocal space nodes along $g_z$ and,
hence, $A \equiv 0$.
Thus, the QEP in Eq.~\ref{eq:qep} reduces to the ordinary eigenvalue problem with respect to $\xi^2$.
Then each eigenstate has two symmetric solutions with $\pm \xi$,
and the near-field is then represented by the series of standing waves, i.e.,
each wave is a sum of two plane waves with an opposite phase.
This reduces the problem to the classical RCWA theory.
To analyze an actual grating, which is not necessarily homogeneous along the vertical direction, with the RCWA, it must be approximated by box-shaped gratings stacked in layers. The dispersion equation can then be solved separately in each layer. Finally, the solutions in each layer can be stitched together into the $E(\bm{r})$ by solving the boundary condition problem (BCP) at each interface between layers. 

\begin{figure}[t]
	\center
	\begin{tikzpicture}[x=0.75pt,y=0.75pt,yscale=-1,xscale=1]

\draw [color={rgb, 255:red, 202; green, 202; blue, 202 }  ,draw opacity=1 ][fill={rgb, 255:red, 252; green, 240; blue, 225 }  ,fill opacity=0.25 ]   (15,90) -- (43,90) -- (105,90) -- (105,75) -- (90,75) -- (75,15) -- (45,15) -- (30,75) -- (15,75) -- cycle ;
\draw [color={rgb, 255:red, 155; green, 155; blue, 155 }  ,draw opacity=1 ][fill={rgb, 255:red, 252; green, 240; blue, 225 }  ,fill opacity=1 ]   (195,180) -- (285,180) -- (285,165) -- (270,165) -- (255,105) -- (225,105) -- (210,165) -- (195,165) -- cycle ;
\draw [color={rgb, 255:red, 202; green, 202; blue, 202 }  ,draw opacity=1 ][fill={rgb, 255:red, 252; green, 240; blue, 225 }  ,fill opacity=0.25 ]   (105,90) -- (133,90) -- (195,90) -- (195,75) -- (180,75) -- (165,15) -- (135,15) -- (120,75) -- (105,75) -- cycle ;
\draw [color={rgb, 255:red, 202; green, 202; blue, 202 }  ,draw opacity=1 ][fill={rgb, 255:red, 252; green, 240; blue, 225 }  ,fill opacity=0.25 ]   (195,90) -- (223,90) -- (285,90) -- (285,75) -- (270,75) -- (255,15) -- (225,15) -- (210,75) -- (195,75) -- cycle ;
\draw [color={rgb, 255:red, 202; green, 202; blue, 202 }  ,draw opacity=1 ][fill={rgb, 255:red, 252; green, 240; blue, 225 }  ,fill opacity=0.25 ]   (105,270) -- (133,270) -- (195,270) -- (195,255) -- (180,255) -- (165,195) -- (135,195) -- (120,255) -- (105,255) -- cycle ;
\draw [color={rgb, 255:red, 202; green, 202; blue, 202 }  ,draw opacity=1 ][fill={rgb, 255:red, 252; green, 240; blue, 225 }  ,fill opacity=0.25 ]   (15,270) -- (43,270) -- (105,270) -- (105,255) -- (90,255) -- (75,195) -- (45,195) -- (30,255) -- (15,255) -- cycle ;
\draw [color={rgb, 255:red, 202; green, 202; blue, 202 }  ,draw opacity=1 ][fill={rgb, 255:red, 252; green, 240; blue, 225 }  ,fill opacity=0.25 ]   (195,270) -- (223,270) -- (285,270) -- (285,255) -- (270,255) -- (255,195) -- (225,195) -- (210,255) -- (195,255) -- cycle ;
\draw    (108,285) -- (192,285) ;
\draw [shift={(195,285)}, rotate = 180] [fill={rgb, 255:red, 0; green, 0; blue, 0 }  ][line width=0.08]  [draw opacity=0] (8.04,-3.86) -- (0,0) -- (8.04,3.86) -- cycle    ;
\draw [shift={(105,285)}, rotate = 0] [fill={rgb, 255:red, 0; green, 0; blue, 0 }  ][line width=0.08]  [draw opacity=0] (8.04,-3.86) -- (0,0) -- (8.04,3.86) -- cycle    ;
\draw [line width=0.75]  [dash pattern={on 9pt off 3.75pt}]  (15,90) -- (285,90) ;
\draw [color={rgb, 255:red, 155; green, 155; blue, 155 }  ,draw opacity=1 ][fill={rgb, 255:red, 252; green, 240; blue, 225 }  ,fill opacity=1 ]   (105,180) -- (195,180) -- (195,165) -- (180,165) -- (165,105) -- (135,105) -- (120,165) -- (105,165) -- cycle ;
\draw [color={rgb, 255:red, 155; green, 155; blue, 155 }  ,draw opacity=1 ][fill={rgb, 255:red, 252; green, 240; blue, 225 }  ,fill opacity=1 ]   (15,180) -- (105,180) -- (105,165) -- (90,165) -- (75,105) -- (45,105) -- (30,165) -- (15,165) -- cycle ;
\draw    (300,93) -- (300,177) ;
\draw [shift={(300,180)}, rotate = 270] [fill={rgb, 255:red, 0; green, 0; blue, 0 }  ][line width=0.08]  [draw opacity=0] (8.04,-3.86) -- (0,0) -- (8.04,3.86) -- cycle    ;
\draw [shift={(300,90)}, rotate = 90] [fill={rgb, 255:red, 0; green, 0; blue, 0 }  ][line width=0.08]  [draw opacity=0] (8.04,-3.86) -- (0,0) -- (8.04,3.86) -- cycle    ;
\draw [line width=0.75]  [dash pattern={on 9pt off 3.75pt}]  (15,180) -- (285,180) ;

\draw (111,294) node [anchor=north west][inner sep=0.75pt]  [font=\small] [align=left] {Periodic BCP};
\draw (308,168) node [anchor=north west][inner sep=0.75pt]  [font=\small] [rotate=-270] [align=left] {Fixed BCP};

\end{tikzpicture}
	\caption{A sketch illustrating the idea from~\cite{huang2013fourier}:
	The lamellar 1D lattice can be represented as a 2D photonic crystal
	that occupies exactly one period in the vertical direction,
	while maintaining periodicity along the lateral direction.}
	\label{fig:idea_sketch}
\end{figure}
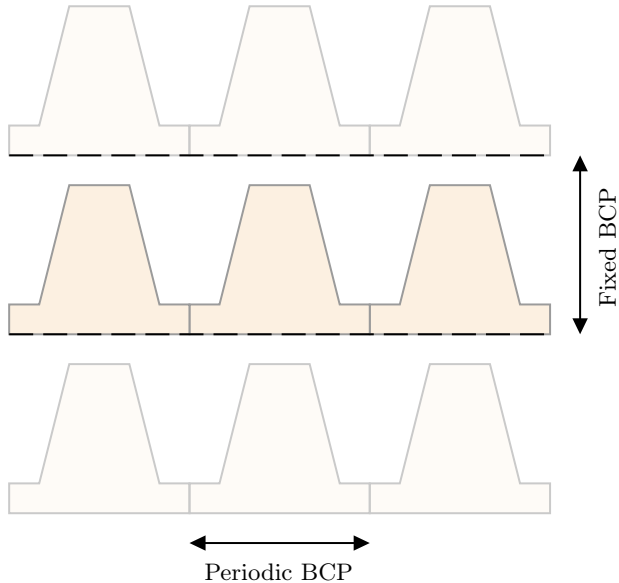

An alternative idea is proposed in~\cite{huang2013fourier}.
Instead of dividing the structure into layers,
one can consider the structure that is periodic in one direction as a 2D periodic structure,
where periodicity in the vertical direction takes up just a single period
(see Fig.~\ref{fig:idea_sketch}).
This allows to take into account the structure change in the vertical direction
by using the 2D Fourier transform (cf. Eq.~\ref{eq:chi}),
so that the BCP has to be solved only for the substrate and the ambient.
However, this violates the $\pm\xi$ symmetry, and one now has to deal with QEP.

Nevertheless, the QEP can be solved by an elegant linearization.
In the context of X-ray diffraction simulations,
this was first shown in~\cite{colella1974multiple}.
Let $D = \xi E$, then Eq.~\ref{eq:qep} becomes $\xi D + AD + BE = 0$,
and these two equations can now be written in a block matrix form
\begin{equation}
	\left[
		\begin{array}{rr}
			0 &  I \\
			-B & -A \\
		\end{array}
		\right]
	\left[
		\begin{array}{r}
			E \\
			D \\
		\end{array}
		\right]
	=
	\xi
	\left[
		\begin{array}{r}
			E \\
			D \\
		\end{array}
	\right]
	.
\end{equation}
Thus the problem is linearized into an eigenvalue problem,
with eigenvectors $[E\; D]^T$ and eigenvalues $\xi$.
The top half of an eigenvector represent an eigenstate,
and amplitudes $\hat E_g$ in Eq.~\ref{eq:bloch} can be calculated as a linear combination of the eigenstates.

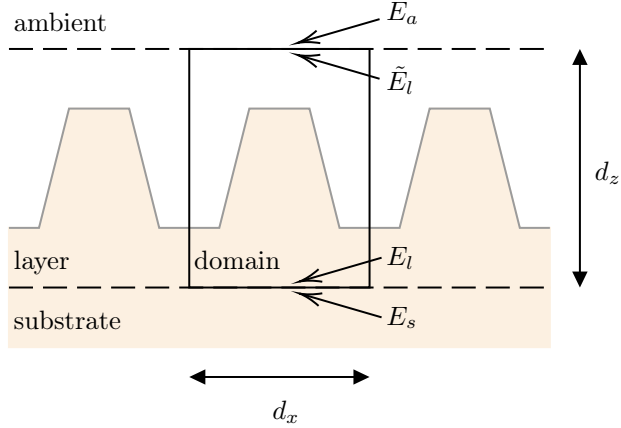
\begin{figure}[t]
	\center
	\begin{tikzpicture}[x=0.75pt,y=0.75pt,yscale=-1,xscale=1]

\draw [color={rgb, 255:red, 252; green, 240; blue, 225 }  ,draw opacity=1 ][fill={rgb, 255:red, 252; green, 240; blue, 225 }  ,fill opacity=1 ]   (30,135) -- (45,135) -- (60,75) -- (90,75) -- (105,135) -- (135,135) -- (150,75) -- (180,75) -- (195,135) -- (225,135) -- (240,75) -- (270,75) -- (285,135) -- (300,135) -- (300,165) -- (30,165) -- cycle ;
\draw [color={rgb, 255:red, 155; green, 155; blue, 155 }  ,draw opacity=1 ]   (30,135) -- (45,135) -- (60,75) -- (90,75) -- (105,135) -- (135,135) -- (150,75) -- (180,75) -- (195,135) -- (225,135) -- (240,75) -- (270,75) -- (285,135) -- (300,135) ;
\draw  [color={rgb, 255:red, 252; green, 240; blue, 225 }  ,draw opacity=1 ][fill={rgb, 255:red, 252; green, 240; blue, 225 }  ,fill opacity=1 ] (30,165) -- (300,165) -- (300,195) -- (30,195) -- cycle ;
\draw [line width=0.75]  [dash pattern={on 9pt off 3.75pt}]  (30,45) -- (300,45) ;
\draw [line width=0.75]  [dash pattern={on 9pt off 3.75pt}]  (30,165) -- (94,165) -- (300,165) ;
\draw   (120,45) -- (210,45) -- (210,165) -- (120,165) -- cycle ;
\draw    (315,48) -- (315,162) ;
\draw [shift={(315,165)}, rotate = 270] [fill={rgb, 255:red, 0; green, 0; blue, 0 }  ][line width=0.08]  [draw opacity=0] (8.04,-3.86) -- (0,0) -- (8.04,3.86) -- cycle    ;
\draw [shift={(315,45)}, rotate = 90] [fill={rgb, 255:red, 0; green, 0; blue, 0 }  ][line width=0.08]  [draw opacity=0] (8.04,-3.86) -- (0,0) -- (8.04,3.86) -- cycle    ;
\draw    (207,210) -- (123,210) ;
\draw [shift={(120,210)}, rotate = 360] [fill={rgb, 255:red, 0; green, 0; blue, 0 }  ][line width=0.08]  [draw opacity=0] (8.04,-3.86) -- (0,0) -- (8.04,3.86) -- cycle    ;
\draw [shift={(210,210)}, rotate = 180] [fill={rgb, 255:red, 0; green, 0; blue, 0 }  ][line width=0.08]  [draw opacity=0] (8.04,-3.86) -- (0,0) -- (8.04,3.86) -- cycle    ;
\draw    (215,30) -- (175.92,41.44) ;
\draw [shift={(174,42)}, rotate = 343.69] [color={rgb, 255:red, 0; green, 0; blue, 0 }  ][line width=0.75]    (10.93,-3.29) .. controls (6.95,-1.4) and (3.31,-0.3) .. (0,0) .. controls (3.31,0.3) and (6.95,1.4) .. (10.93,3.29)   ;
\draw    (215,60) -- (175.92,48.56) ;
\draw [shift={(174,48)}, rotate = 16.31] [color={rgb, 255:red, 0; green, 0; blue, 0 }  ][line width=0.75]    (10.93,-3.29) .. controls (6.95,-1.4) and (3.31,-0.3) .. (0,0) .. controls (3.31,0.3) and (6.95,1.4) .. (10.93,3.29)   ;
\draw    (216,150) -- (176.92,161.44) ;
\draw [shift={(175,162)}, rotate = 343.69] [color={rgb, 255:red, 0; green, 0; blue, 0 }  ][line width=0.75]    (10.93,-3.29) .. controls (6.95,-1.4) and (3.31,-0.3) .. (0,0) .. controls (3.31,0.3) and (6.95,1.4) .. (10.93,3.29)   ;
\draw    (216,180) -- (176.92,168.56) ;
\draw [shift={(175,168)}, rotate = 16.31] [color={rgb, 255:red, 0; green, 0; blue, 0 }  ][line width=0.75]    (10.93,-3.29) .. controls (6.95,-1.4) and (3.31,-0.3) .. (0,0) .. controls (3.31,0.3) and (6.95,1.4) .. (10.93,3.29)   ;

\draw (31,25) node [anchor=north west][inner sep=0.75pt]   [align=left] {{ambient}};
\draw (31,145) node [anchor=north west][inner sep=0.75pt]   [align=left] {{layer}};
\draw (31,175) node [anchor=north west][inner sep=0.75pt]   [align=left] {{substrate}};
\draw (217,20) node [anchor=north west][inner sep=0.75pt]    {$E_{a}$};
\draw (217,52) node [anchor=north west][inner sep=0.75pt]    {$\tilde{E_{l}}$};
\draw (217,143) node [anchor=north west][inner sep=0.75pt]    {$E_{l}$};
\draw (217,173) node [anchor=north west][inner sep=0.75pt]    {$E_{s}$};
\draw (121,145) node [anchor=north west][inner sep=0.75pt]   [align=left] {{domain}};
\draw (321,100) node [anchor=north west][inner sep=0.75pt]    {$d_{z}$};
\draw (160,220.4) node [anchor=north west][inner sep=0.75pt]    {$d_{x}$};

\end{tikzpicture}
	\caption{A sketch illustrating the structure of the BCP.
                The system is divided into three parts:
                ambient, structured layer, and substrate.
                The arrows indicate which part the amplitude vectors $E$ refer to.
                The near field is sought within a computational domain.
                }
	\label{fig:bamps}
\end{figure}

The next step is to solve the BCP.
Unlike the RCWA, the solution is quite simple, since there are only two interfaces to consider. There is an interface between the ambient and the structured layer, and between the layer and the substrate. To do so, we consider four amplitude vectors: $E_a$ is the representation of the field in the ambient directly at the surface of the structure, $\tilde E_l$ is also for the field at the surface but inside the structured layer. Correspondingly, $E_l$ is the field at the bottom of the structure and $E_s$ is in the substrate (see Fig.~\ref{fig:bamps}).
The BCP is solved by imposing the continuity on the amplitudes and their gradients along the vertical direction at the interfaces.
This can be expressed as
\begin{equation}
	M_a E_a
	=
	M_l \tilde E_l 
        ;
	\quad
	M_l E_l
	=
	M_s E_s
	,
	\label{eq:bdc}
\end{equation}
where $M_a$ and $M_s$ describe the refraction in the homogeneous media,
that is the ambient and the substrate, respectively, and $M_l$ corresponds to the structured layer. 

The matrix $M$ has the simplest form if the following order of the components of the amplitude vector in homogeneous media is chosen: the amplitude vectors are divided into two blocks $E = [T\; R]^T$, where $R$ consists of the amplitudes of upward propagating plane waves (decay modes with respect to the direction of $z$) and $T$ is for downward propagating waves (gain modes). Moreover, the order within $R$ and $T$ is such that $n$-th elements of each form a standing wave with an opposite phase:
$T_n e^{-ik_{zn}}+R_n e^{+ik_{zn}}$. Then, for $M$ we can write 
\begin{equation}
	M_{a,s} = \left[
		\begin{array}{rr}
			I &  I \\
			-K & K \\
		\end{array}
		\right],
        \label{eq:homogeneous}
\end{equation}
where
$K = \text{diag}(\ldots \kappa_{0z} \ldots)$.
The vertical components $\kappa_{0z}$ are calculated using the spherical dispersion Eq.~\ref{eq:spherical},
choosing appropriate dielectric constants $\chi_0$ for each matrix (ambient and substrate).
The upper half of a matrix corresponds to the continuity of the field itself,
and the lower half to its gradient,
which for the plane waves simply scales the field with the wave number, hence Eq.~\ref{eq:homogeneous}. 

As described above in the structured medium, the amplitude vectors are the linear combinations of the eigenstates, i.e. the eigenvectors $E$, computed by solving QEP. Let us bundle the eigenvalues $\xi$ into a vector $\Xi$ and the eigenvectors into a matrix $F$. Note that we consider the eigenvectors as columns. 
Then for the matrix $M_l$ we derive:
\begin{equation}
	M_l = \left[
		\begin{array}{c}
			F \\
			(G \oplus k\Xi) \circ F \\
		\end{array}
	 	\right].
	\label{eq:propagation_matrix}
\end{equation}
Here, symbol $\oplus$ denotes the outer sum $(a \oplus b)_{ij} = a_i+b_j$,
and $G$ is the vector composed of vertical coordinates $g_z$ of nodes in the reciprocal space.
Again, the upper half of the matrix $M_l$ corresponds to the field continuity
and the lower half to the gradient continuity.
The matrix $G \oplus k\Xi$ is composed of the vertical components of the wave vectors and, unlike the homogeneous case, is not diagonal, which takes multiple wave interference into account.
The symbol $\circ$ denotes the Hadamar (element-vise) matrix multiplication.

The next step in solving the BCP is to link the amplitudes at the surface of the structured layer and at the bottom. This is done using a unitary matrix that takes into account the phase change and absorption of the propagating waves:
\begin{equation}
        \tilde E_l = Q E_l,
        \label{eq:propagation}
\end{equation}
where
$Q=\text{diag}(e^{ -i k \Xi d_z })$.
To derive the far-field solution, one can solve the Eqs.~\ref{eq:bdc} and~\ref{eq:propagation} with respect to $R_a$ (the second block of $E_a$). The resulting elements of $R_a$ represent the diffraction amplitudes with the corresponding order.
The near-field solution can be obtained by solving the equations with respect to $E_l$ and subsequently applying the Fourier transform to obtain the spatial profile of the electric field, $E(\bm r)$.
For that, one can assume that
$T_a = (\ldots0,1,0\ldots)$,
which corresponds to the incident plane wave:
amplitudes of all higher order wave modes are zero.
Further, one assumes $R_s = (\ldots0,0,0\ldots)$
as the reflection from the bottom of the substrate is negligible.
Thus, from a mathematical standpoint, the problem is solved.

Unfortunately, it is practically impossible to find a numerical solution in this manner.
This is because the matrix $Q$ is ill conditioned as it contains both exponentially increasing and decaying elements.
By back propagating amplitudes from values below the machine epsilon,
one simply calculates computation noise instead of the amplitude values.
To mitigate this problem one has to find a form of Eqs.~\ref{eq:bdc} and~\ref{eq:propagation} including only decaying exponents.
This idea is known and widely applied.
For instance, in the context of the X-ray diffraction it is discussed 
in~\cite{stepanov1998dynamical}.

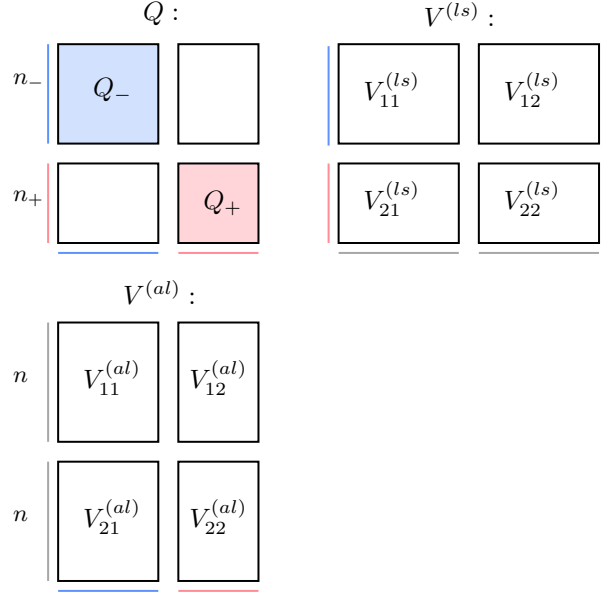
\begin{figure}[t]
	\center
	\begin{tikzpicture}[x=0.75pt,y=0.75pt,yscale=-1,xscale=1]

\draw   (25,170) -- (75,170) -- (75,230) -- (25,230) -- cycle ;
\draw   (85,170) -- (125,170) -- (125,230) -- (85,230) -- cycle ;
\draw   (25,240) -- (75,240) -- (75,300) -- (25,300) -- cycle ;
\draw   (85,240) -- (125,240) -- (125,300) -- (85,300) -- cycle ;
\draw  [fill={rgb, 255:red, 210; green, 225; blue, 254 }  ,fill opacity=1 ] (25,30) -- (75,30) -- (75,80) -- (25,80) -- cycle ;
\draw  [fill={rgb, 255:red, 255; green, 213; blue, 218 }  ,fill opacity=1 ][line width=0.75]  (85,90) -- (125,90) -- (125,130) -- (85,130) -- cycle ;
\draw   (25,90) -- (75,90) -- (75,130) -- (25,130) -- cycle ;
\draw   (85,30) -- (125,30) -- (125,80) -- (85,80) -- cycle ;
\draw   (165,30) -- (225,30) -- (225,80) -- (165,80) -- cycle ;
\draw   (235,30) -- (295,30) -- (295,80) -- (235,80) -- cycle ;
\draw   (165,90) -- (225,90) -- (225,130) -- (165,130) -- cycle ;
\draw   (235,90) -- (295,90) -- (295,130) -- (235,130) -- cycle ;
\draw [color={rgb, 255:red, 94; green, 147; blue, 254 }  ,draw opacity=1 ][line width=0.75]    (20,30) -- (20,80) ;
\draw [color={rgb, 255:red, 255; green, 140; blue, 151 }  ,draw opacity=1 ]   (20,90) -- (20,130) ;
\draw [color={rgb, 255:red, 167; green, 167; blue, 167 }  ,draw opacity=1 ]   (20,170) -- (20,230) ;
\draw [color={rgb, 255:red, 167; green, 167; blue, 167 }  ,draw opacity=1 ]   (20,240) -- (20,300) ;
\draw [color={rgb, 255:red, 167; green, 167; blue, 167 }  ,draw opacity=1 ]   (165,135) -- (225,135) ;
\draw [color={rgb, 255:red, 167; green, 167; blue, 167 }  ,draw opacity=1 ]   (235,135) -- (295,135) ;
\draw [color={rgb, 255:red, 255; green, 140; blue, 151 }  ,draw opacity=1 ]   (160,90) -- (160,130) ;
\draw [color={rgb, 255:red, 94; green, 147; blue, 254 }  ,draw opacity=1 ][line width=0.75]    (160,31) -- (160,81) ;
\draw [color={rgb, 255:red, 94; green, 147; blue, 254 }  ,draw opacity=1 ][line width=0.75]    (25,305) -- (75,305) ;
\draw [color={rgb, 255:red, 255; green, 140; blue, 151 }  ,draw opacity=1 ]   (125,305) -- (85,305) ;
\draw [color={rgb, 255:red, 94; green, 147; blue, 254 }  ,draw opacity=1 ][line width=0.75]    (25,135) -- (75,135) ;
\draw [color={rgb, 255:red, 255; green, 140; blue, 151 }  ,draw opacity=1 ]   (125,135) -- (85,135) ;

\draw (41,45.4) node [anchor=north west][inner sep=0.75pt]    {$Q_{-}$};
\draw (96,102.4) node [anchor=north west][inner sep=0.75pt]    {$Q_{+}$};
\draw (36,187.4) node [anchor=north west][inner sep=0.75pt]    {$V_{11}^{( al)}$};
\draw (89,187.4) node [anchor=north west][inner sep=0.75pt]    {$V_{12}^{( al)}$};
\draw (36,257.4) node [anchor=north west][inner sep=0.75pt]    {$V_{21}^{( al)}$};
\draw (89,257.4) node [anchor=north west][inner sep=0.75pt]    {$V_{22}^{( al)}$};
\draw (176,42.4) node [anchor=north west][inner sep=0.75pt]    {$V_{11}^{( ls)}$};
\draw (246,42.4) node [anchor=north west][inner sep=0.75pt]    {$V_{12}^{( ls)}$};
\draw (246,97.4) node [anchor=north west][inner sep=0.75pt]    {$V_{22}^{( ls)}$};
\draw (176,97.4) node [anchor=north west][inner sep=0.75pt]    {$V_{21}^{( ls)}$};
\draw (66,7.4) node [anchor=north west][inner sep=0.75pt]    {$Q:$};
\draw (56,147.4) node [anchor=north west][inner sep=0.75pt]    {$V^{( al)} :$};
\draw (206,7.4) node [anchor=north west][inner sep=0.75pt]    {$V^{( ls)} :$};
\draw (1,43.4) node [anchor=north west][inner sep=0.75pt]  [font=\small]  {$n_{-}$};
\draw (1,102.4) node [anchor=north west][inner sep=0.75pt]  [font=\small]  {$n_{+}$};
\draw (1,193.4) node [anchor=north west][inner sep=0.75pt]  [font=\small]  {$n$};
\draw (1,263.4) node [anchor=north west][inner sep=0.75pt]  [font=\small]  {$n$};

\end{tikzpicture}
	\caption{Block matrix diagram for Eq.~\ref{eq:bdc_new}.
	The size of the matrices is shown schematically.
	Here $N$ is the number of wave modes considered in the ansatz,
	$n_+$ is the number of amplification modes,
	$n_-$ is the number of decay modes.
	}
	\label{fig:bdc_new}
\end{figure}

Let us now implement this idea for our problem.
Generally, the order at which the eigenvalues $\xi$ are sorted is arbitrary,
but one need to separate gain and decay wave modes.
Therefore we sort and split the vector $\xi$ of eigenvalues into a block vector
$\Xi = [\Xi_- \; \Xi_+]^T$,
such that $\Xi_-$ contains eigenvalues with
$\text{Im}\, \xi < 0$
(gain modes) and $\Xi_+$ contains
$\text{Im}\, \xi > 0$ (decay modes).
Let us assume that the number of decay modes is not the same as the number of gain modes.
Such an assumption is incorrect because the dispersion is described by a quadratic equation.
However, it can occur numerically. 
The sequence of column eigenvectors in $F$ must be sorted accordingly.
Furthermore, one should rewrite $Q$ as a block matrix as well:
\begin{equation}
	Q = \left[
		\begin{array}{cc}
			Q_- &  0 \\
			0 &  Q_+ \\
		\end{array}
	\right]
	.
\end{equation}
Now we can rewrite Eqs.~\ref{eq:bdc} and~\ref{eq:propagation} in the block matrix form
%
\begin{equation}
	\left[
		\begin{array}{c}
			T_l \\
			R_a \\
		\end{array}
		\right]
	=
	P_{al}
	\left[
		\begin{array}{c}
			T_a \\
			R_l \\
		\end{array}
		\right];
	\quad
	\left[
		\begin{array}{c}
			T_s \\
			R_l \\
		\end{array}
		\right]
	=
	P_{ls}
	\left[
		\begin{array}{c}
			T_l \\
			R_s \\
		\end{array}
	\right]
	.
	\label{eq:bdc_new}
\end{equation}
Notice that here the order of propagation is changed:
each amplitude vector contains amplitudes from different media.
Now, using block-matrix algebra,
one can easily find an explicit expression for both
$P_{al}$ and $P_{ls}$:
\begin{equation}
	P =
	\left[
		\begin{array}{cc}
			Q_-^*V_{11}^\dagger & -Q_-^*V_{11}^\dagger V_{12} Q_+ \\
			V_{21} V_{11}^\dagger & V_{22}Q_+ - V_{21} V_{11}^\dagger V_{12} Q_+ \\
		\end{array}
	\right]
	.
\end{equation}
Here, the "$\dagger$" superscript denotes the Moore-Penrose inverse (or pseudoinverse),
it is a generalization of inversion for non-square matrices;
$V$ is the $2\times 2$ block matrix with blocks $V_{ij}$.
For $P_{al}$, one has to take $V^{(al)} = M_a^{-1} M_l $.
For $P_{ls}$, one has to take $V^{(ls)} = M_l^\dagger M_s$ and to set $Q = I$,
this is because we do not consider the propagation inside the substrate.
Again, this is because we no longer rely on there being the same number of decay modes as there are gain modes.
Therefore, the sizes of the blocks in Eq.~\ref{eq:bdc_new} are different.
See the Fig.~\ref{fig:bdc_new} for an illustration.
Note that the matrix $P$ does not explicitly contain $Q_-$.
It contains only its inversion, which,
being a unitary matrix, can be computed as $Q_-^{-1} = Q_-^*$. 
Now the $Q_-^*$ matrix also contains decaying exponents.
This eliminates the problem of backpropagating values below the machine epsilon.

%

Now the amplitude coefficients can be determined as
\begin{equation}
	T_l = (I - A^{(al)}_{12}A^{(gs)}_{21} )^{-1} A^{(al)}_{11} T_a
        ,
\end{equation}
and
\begin{equation}
	R_l = (I - A^{(ls)}_{21}A^{(al)}_{12} )^{-1} A^{(ls)}_{21} A^{(al)}_{11} T_a
        .
\end{equation}
Having computed the vectors $T_l$ and $R_l$, one can find $E(\bm r)$
simply by applying the fast Fourier transform to solve Eq.~\ref{eq:bloch}.
Thus, the near-field problem is solved.
In this framework one solves the far-field problem by finding the transmission $T_s$
and reflection $R_a$ amplitudes:
\begin{equation}
	T_s = A^{(ls)}_{11} T_l,
	\quad
	R_a = A^{(al)}_{21} T_a + A^{(al)}_{22} R_l.
	\label{eq:end}
\end{equation}
By obtaining these amplitude vector the problem is solved.

The last theoretical consideration of this work is the model of the structure itself.
To solve the near-field problem
one needs to represent the structure in terms of the Fourier transform in Eq.~\ref{eq:chi}.
We propose to approximate the grating profile with a polygon.
The Fourier transform of the indicator function $\gamma$ of a polygon
($\gamma(\bm r) = 1$ if $\bm r \in \Gamma$  and $\gamma(\bm r) = 0$ otherwise)
have a simple closed form solution.
It was first mentioned in~\cite{shung_wu1983fourier}.
In~\cite{wuttke2021numerically} the numerically stable solution was proposed to eliminate divergence near $Q = 0$.
However, we do not want to deal with such problems since we consider our structure to be periodic,
so we only compute the transform of $\chi$ on a discrete grid,
so our only problem is to get rid of the singularity at exactly $Q = 0$.

%
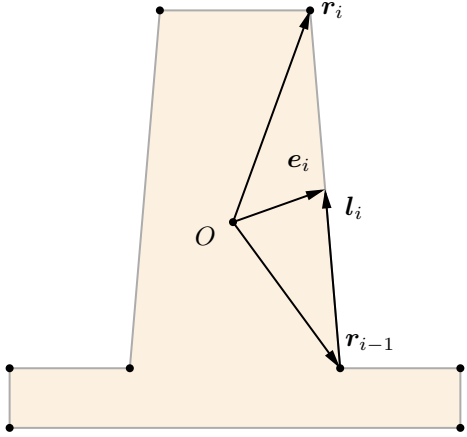
\begin{figure}[t]
	\center
	\tikzset{every picture/.style={line width=0.75pt}} 

\begin{tikzpicture}[x=0.75pt,y=0.75pt,yscale=-1,xscale=1]

\draw [color={rgb, 255:red, 209; green, 209; blue, 209 }  ,draw opacity=1 ]   (210,240) -- (156.5,166.5) ;
\draw [color={rgb, 255:red, 209; green, 209; blue, 209 }  ,draw opacity=1 ]   (195,60) -- (156.5,166.5) ;
\draw [color={rgb, 255:red, 172; green, 172; blue, 172 }  ,draw opacity=1 ][fill={rgb, 255:red, 252; green, 240; blue, 225 }  ,fill opacity=1 ]   (45,270) -- (270,270) -- (270,240) -- (210,240) -- (195,60) -- (120,60) -- (105,240) -- (45,240) -- cycle ;
\draw  [fill={rgb, 255:red, 0; green, 0; blue, 0 }  ,fill opacity=1 ] (118.5,60) .. controls (118.5,59.17) and (119.17,58.5) .. (120,58.5) .. controls (120.83,58.5) and (121.5,59.17) .. (121.5,60) .. controls (121.5,60.83) and (120.83,61.5) .. (120,61.5) .. controls (119.17,61.5) and (118.5,60.83) .. (118.5,60) -- cycle ;
\draw  [fill={rgb, 255:red, 0; green, 0; blue, 0 }  ,fill opacity=1 ] (193.5,60) .. controls (193.5,59.17) and (194.17,58.5) .. (195,58.5) .. controls (195.83,58.5) and (196.5,59.17) .. (196.5,60) .. controls (196.5,60.83) and (195.83,61.5) .. (195,61.5) .. controls (194.17,61.5) and (193.5,60.83) .. (193.5,60) -- cycle ;
\draw  [fill={rgb, 255:red, 0; green, 0; blue, 0 }  ,fill opacity=1 ] (208.5,240) .. controls (208.5,239.17) and (209.17,238.5) .. (210,238.5) .. controls (210.83,238.5) and (211.5,239.17) .. (211.5,240) .. controls (211.5,240.83) and (210.83,241.5) .. (210,241.5) .. controls (209.17,241.5) and (208.5,240.83) .. (208.5,240) -- cycle ;
\draw  [fill={rgb, 255:red, 0; green, 0; blue, 0 }  ,fill opacity=1 ] (268.5,240) .. controls (268.5,239.17) and (269.17,238.5) .. (270,238.5) .. controls (270.83,238.5) and (271.5,239.17) .. (271.5,240) .. controls (271.5,240.83) and (270.83,241.5) .. (270,241.5) .. controls (269.17,241.5) and (268.5,240.83) .. (268.5,240) -- cycle ;
\draw  [fill={rgb, 255:red, 0; green, 0; blue, 0 }  ,fill opacity=1 ] (268.5,270) .. controls (268.5,269.17) and (269.17,268.5) .. (270,268.5) .. controls (270.83,268.5) and (271.5,269.17) .. (271.5,270) .. controls (271.5,270.83) and (270.83,271.5) .. (270,271.5) .. controls (269.17,271.5) and (268.5,270.83) .. (268.5,270) -- cycle ;
\draw  [fill={rgb, 255:red, 0; green, 0; blue, 0 }  ,fill opacity=1 ] (43.5,270) .. controls (43.5,269.17) and (44.17,268.5) .. (45,268.5) .. controls (45.83,268.5) and (46.5,269.17) .. (46.5,270) .. controls (46.5,270.83) and (45.83,271.5) .. (45,271.5) .. controls (44.17,271.5) and (43.5,270.83) .. (43.5,270) -- cycle ;
\draw  [fill={rgb, 255:red, 0; green, 0; blue, 0 }  ,fill opacity=1 ] (43.5,240) .. controls (43.5,239.17) and (44.17,238.5) .. (45,238.5) .. controls (45.83,238.5) and (46.5,239.17) .. (46.5,240) .. controls (46.5,240.83) and (45.83,241.5) .. (45,241.5) .. controls (44.17,241.5) and (43.5,240.83) .. (43.5,240) -- cycle ;
\draw  [fill={rgb, 255:red, 0; green, 0; blue, 0 }  ,fill opacity=1 ] (103.5,240) .. controls (103.5,239.17) and (104.17,238.5) .. (105,238.5) .. controls (105.83,238.5) and (106.5,239.17) .. (106.5,240) .. controls (106.5,240.83) and (105.83,241.5) .. (105,241.5) .. controls (104.17,241.5) and (103.5,240.83) .. (103.5,240) -- cycle ;
\draw    (210,240) -- (202.67,151.99) ;
\draw [shift={(202.5,150)}, rotate = 85.24] [fill={rgb, 255:red, 0; green, 0; blue, 0 }  ][line width=0.08]  [draw opacity=0] (9.6,-2.4) -- (0,0) -- (9.6,2.4) -- cycle    ;
\draw    (156.5,166.5) -- (200.62,150.68) ;
\draw [shift={(202.5,150)}, rotate = 160.27] [fill={rgb, 255:red, 0; green, 0; blue, 0 }  ][line width=0.08]  [draw opacity=0] (9.6,-2.4) -- (0,0) -- (9.6,2.4) -- cycle    ;
\draw    (156.5,166.5) -- (194.32,61.88) ;
\draw [shift={(195,60)}, rotate = 109.88] [fill={rgb, 255:red, 0; green, 0; blue, 0 }  ][line width=0.08]  [draw opacity=0] (9.6,-2.4) -- (0,0) -- (9.6,2.4) -- cycle    ;
\draw    (156.5,166.5) -- (208.82,238.38) ;
\draw [shift={(210,240)}, rotate = 233.95] [fill={rgb, 255:red, 0; green, 0; blue, 0 }  ][line width=0.08]  [draw opacity=0] (9.6,-2.4) -- (0,0) -- (9.6,2.4) -- cycle    ;
\draw  [fill={rgb, 255:red, 0; green, 0; blue, 0 }  ,fill opacity=1 ] (155,166.5) .. controls (155,165.67) and (155.67,165) .. (156.5,165) .. controls (157.33,165) and (158,165.67) .. (158,166.5) .. controls (158,167.33) and (157.33,168) .. (156.5,168) .. controls (155.67,168) and (155,167.33) .. (155,166.5) -- cycle ;

\draw (211,221.4) node [anchor=north west][inner sep=0.75pt]    {$\bm{r}_{i-1}$};
\draw (199,54.4) node [anchor=north west][inner sep=0.75pt]    {$\bm{r}_{i}$};
\draw (182,131.4) node [anchor=north west][inner sep=0.75pt]    {$\bm{e}_{i}$};
\draw (211,152.4) node [anchor=north west][inner sep=0.75pt]    {$\bm{l}_{i}$};
\draw (136,167.4) node [anchor=north west][inner sep=0.75pt]    {$O$};

\end{tikzpicture}
	\caption{A sketch of a non-self-intersecting polygon $\Gamma$.
		The origin $O$ can be chosen arbitrarily.
		The vertices $\bm{r}_i$ are indexed counterclockwise.
		The vectors $\bm{e}_i$ and $\bm{l}_i$ are involved in Eq.~\ref{eq:poly}.
		}

	\label{fig:poly}
\end{figure}

Consider a non-self-intersecting polygon $\Gamma$ with the indicator $\gamma$.
By definition, the Fourier transform of $\gamma$ is:
\begin{equation}
	\hat\gamma(\bm q) = \int_\Gamma d^2r e^{-i\bm{qr}}
	.
	\label{eq:form_factor}
\end{equation}
Let a polygon to be described with a set of vertices with coordinates $\{\bm{r}_i\}$
(see Fig.~\ref{fig:poly}).
The set $\{\bm{r}_i\}$ the set is ordered,
such that the vertices are in the counter clockwise sequence.
By the virtue of the Stokes theorem,
the integration in Eq.~\ref{eq:form_factor}
can be turned int the integration over the line segments of the polygon $\Gamma$.
See detailed proof in~\cite{wuttke2021numerically}.
The integration over the line segments yields:
\begin{equation}
	\hat\gamma (\bm{q})
	=
	\sum_j \dfrac{q_y l_{yj} - q_x l_{xj}}{q}
	\left\{
		\text{sinc} (-\bm{q} \bm{l}_j)
		e^{-i \bm{q}\bm{e}_j }
		-1
	\right\}
	,
	\label{eq:poly}
\end{equation}
with $\bm{e}_j = (\bm{r}_j +  \bm{r}_{j-1}) / 2$ and
$\bm{l}_j = (\bm{r}_j -  \bm{r}_{j-1}) / 2$.

The described theoretical considerations constitute the polygonal approach.
Here is the numeric recipe for calculating the diffraction on the grating.
Start by calculating the dielectric susceptibility constant $\chi$ for the material 
for a given incident of photon energy using tabular data.
Create a polygon model of the grating's line profile
by specifying the coordinates of each vertex
and putting them into the sequence in counterclockwise order.
Calculate the Fourier components $\chi_g$ by calculating the $\chi\hat\gamma( \bm{g} )$ at all considered reciprocal space nodes $\{\bm{g}\}$, where $\chi$ is the dielectric constant of the material.
Further, put these components $\chi_g$ in the Toeplitz matrix Eq.~\ref{eq:toeplitz}.
Note that the components $\chi_g$ are calculated on a two-dimensional grid,
but to compose a Toeplitz matrix, one has to reorder them into a one-dimensional array.
The order of that array is arbitrary, but it will define the order in the amplitude vectors $R$ and $T$.
Solve the linearized QEP Eq.~\ref{eq:qep} using any appropriate eigenvalue solver.
Sort the eigenvalues into two sets:
${\rm Im} \, \xi > 0$ -- decay modes
and ${\rm Im} \, \xi < 0$ -- gain modes.
Sort column vectors in the eigenmatrix accordingly.
The upper half block of the eigenvector matrix is the matrix $F$.
Using this matrix, compute the propagation matrix $M_l$ in Eq.~\ref{eq:propagation_matrix}.
Using Eqs.~\ref{eq:bdc_new}-\ref{eq:end} compute amplitude vectors.
Amplitude $R_a$ describes the far-field of the scattered wave.
To compute near-field solve Eq.~\ref{eq:bloch} through an FFT algorithm.

\section{Numeric simulations and data analysis}

To test the feasibility of the polygonal approach,
we used it for simulations of X-ray scattering from nanoscale lamellar gratings.
We test the approach in two ways.
First, we compared its numeric simulations with simulations carried out by other existing approaches.
Second, we compared the polygonal approach results to the experimental data.
Both are performed in the context of grazing-incidence small-angle X-ray scattering (GISAXS).
In the GISAXS experiment,
the sample is irradiated with a monochromatic X-ray beam under the grazing incidence conditions
in which the angle of incidence is chosen near the critical angle (1 to 5 critical angles).
For all further analysis, we consider conical mount, i.e.,
the azimuthal orientation of a grating with sidewalls parallel to the incident beam
(azimuth angle $\phi = 0$).
In such an experimental setting, one measures the small-angle scattering pattern,
in which bright diffraction peaks and lower-intensity diffuse scattering effects can be observed.
Diffuse scattering is due to the correlation in structural imperfections;
it is out of the scope of this study.
Nevertheless, the diffraction pattern encodes the averaged periodic structure of the sample.
The reconstruction of the structure is of interest in the context of nano-metrology research.
We use our approach to calculate the diffraction in the geometry described above,
which is called the conical diffraction.
The integrated intensities of the diffraction peaks which should be proportional to the $R$ amplitudes of the solution
in the ambient above the sample provide the far-field or k-representation of the solution.
The same solution can also be represented as the $E(\bm{r})$ calculated near the sample,
which is the near-field or r-representation.

\begin{figure*}[t]
	\center
	\includegraphics[width = 1\textwidth]{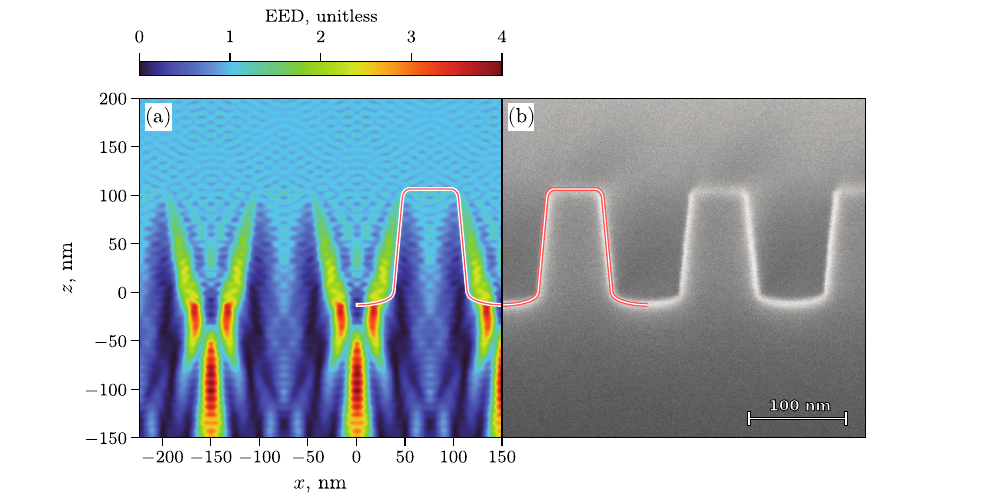}
	\caption{The near-field EED map (a)
	calculated for the best-fit model,
	represented by the solid line,
        and the SEM image (a) of the silicon grating sample.
	}
	\label{fig:tem_nf}
\end{figure*}

We performed the near-field computation using the conventional RCWA, the FEM, and the polygonal approach.
For the nano structured surface, an idealized model of a lamellar silicon grating is taken,
with the lateral period (the pitch) $D=150$~nm.
The line height of the grating is $D=120$~nm.
The sidewalls of the sample model are inclined with an angle $\tau$, $\tan\tau \approx 0.1$.
The surface is irradiated by a monochromatic beam with the photon energy
$\hbar\omega = 5500$~eV under the grazing incidence angle $\alpha_i = 0.5^\circ$ in conical scattering geometry.
The dielectric susceptibility $\chi$ is calculated for the pure silicon
with nominal density $\rho = 2.33$~g/cm$^3$ according to the tabular data from~\cite{henke1993xray}.
The parameters are chosen to be comparable with the actual sample (see its SEM image in Fig.~\ref{fig:tem_nf}a) which we discuss later in the text.

\begin{figure*}[t]
	\center
	\includegraphics[width = 1\textwidth]{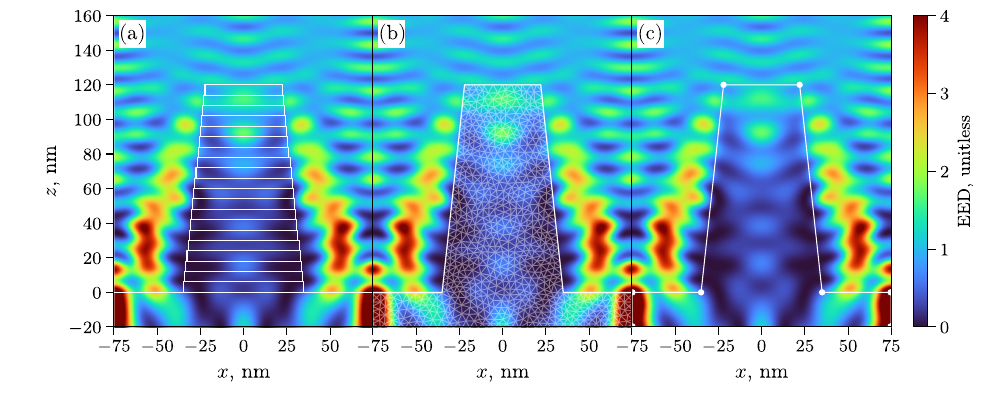}
	\caption{Near field calculation comparison. Electric energy density $|E|^2$ (EED) calculated by
	(a) RCWA,
	(b) FEM and
	(c) polygonal approach.
	Test model:
	silicon grating with $150$~nm lateral period,
	incident beam photon energy $\hbar\omega = 5500$~eV,
	angle of incidence $\alpha_i = 0.5^\circ$,
	conical mount ($\phi = 0^\circ$).}
	\label{fig:nf}
\end{figure*}

A visual comparison of the three computations are shown in Fig.~\ref{fig:nf}.
The near-field solutions are represented as the normalized electric energy density (EED) $|E(r)|^2$.
For the RCWA computation, the grating line is divided into 20 layers,
and the number of Fourier modes taken for the ansatz is 41.
These numbers of layers and modes appear to be sufficient, according to a convergence study.
The ambient and the substrate are considered semi-infinite laterally homogeneous layers.
The structure subdivision into the layers is schematically shown over the EED map in Fig.~\ref{fig:nf}(a).

The FEM computation is done using the JCMWave software~\cite{pomplun2007adaptive}.
This package is often used as a reference for nano-metrology
as it is widely used in the relevant literature (see among others~\cite{tockhorn2022nano,vaskin2019manipulation,soltwisch2018element})
and has also been optimized for the very short X-ray wavelengths.
Hence, we can consider the FEM computation as a benchmark.
In the FEM, the computational domain is meshed into triangular elements or patches.
In each patch, an approximation to the solution is sought as a linear combination of polynomial basis functions.
This allows to handle structures with complex shapes of line profiles.
The finite element side boundary of $h = 6$~nm and a polynomial degree of $p = 4$ were used in the computation.
The numerical precision in the simulation was verified until the quasi-exact result was achieved.
The FEM results are shown in Fig.~\ref{fig:nf}(b), together with the schematic representation of the computation mesh.
The elements representing the ambient part of the computational domain are omitted for visual clarity.

Finally, we perform the computation with the polygonal approach using the numeric description given above.
The computational domain of the same size as for the FEM computation is chosen,
where a polygon with eight vertices represents the line and part of the substrate.
In terms of nodes in reciprocal space,
the computational domain size was chosen according to a convergence study.
Computational accuracy increases with the number of nodes in the reciprocal space at the expense of computation time.
Therefore, one aims to minimize the number of nodes while maintaining sufficient accuracy.
To do this, we use the convergence study:
the number of nodes is increased until the $L_2$ norm of the solution is saturated with a given accuracy.
The size of $41\times21$ nodes appears to be sufficient for this calculation.
The results are shown in Fig.~\ref{fig:nf}(c).

By visual inspection, the results are in excellent qualitative agreement.
Let us use the FEM result as the reference.
To compare the quality of results, we calculate the relative $L_2$ norms
\begin{equation*}
	\dfrac{ || \, |E|-|E_{\rm FEM}| \, ||_2 }{ || \,|E_{\rm FEM}|\, ||_2 }
	=
	\dfrac{\sqrt{\sum|\, |E|-|E_{\rm FEM}| \,|^2}}{\sqrt{\sum |E_{\rm FEM}|^2}}
	.
\end{equation*}
Absolute values here compensate for an arbitrary choice of the incident beam phase.
The sum is taken over the points of arrays $E$.
We use the FEM solution as a reference.
Such norm for the difference between RCWA and FEM is
$0.01$,
and between the polygonal approach and FEM is $0.02$.
Therefore, the polygonal approach provides an approximate solution of the same order of accuracy as RCWA and FEM.

\begin{figure}[t]
	\center
	\includegraphics[width = .5\textwidth]{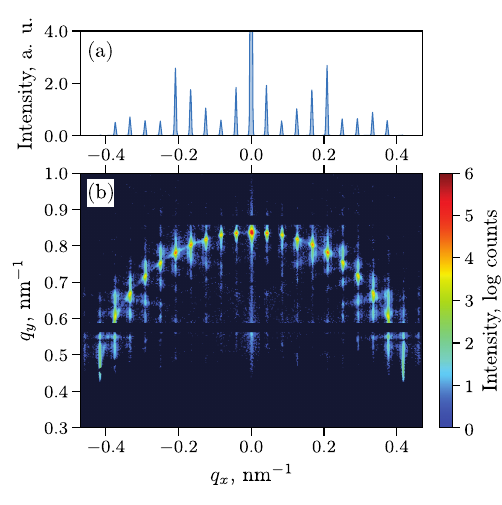}
	\caption{The GISAXS data of the silicon lamellar lattice originally published in~\cite{soltwisch2017reconstructing}.
	Incident beam photon energy $\hbar\omega = 5500$~eV,
	angle of incidence $\alpha_i = 0.86^\circ$, conical mount.
	The scattering pattern (b) is shown on a logarithmic scale.
	The line cut along the diffraction cone (a) is with respect to the reciprocal coordinate $q_x$.
	Integrals of the peaks in (a) are proportional to the corresponding diffraction efficiencies.}
	\label{fig:pattern}
\end{figure}

Furthermore, we compare the polygonal approach simulations with the conical diffraction data measured in the GISAXS geometry,
taken from a previously published research~\cite{soltwisch2017reconstructing}.
In~\cite{soltwisch2017reconstructing}
the structure parameters of an actual nanoscale grating were reconstructed by fitting the simulation to the data.

A fixed set of geometric parameters,
such as height, sidewall angle, pitch, etc.,
is used to describe the structure profile.
Then, such a model was used for the forward diffraction simulation by the FEM.
By repeating this process iteratively employing optimization algorithms,
the best-fit set of parameters was found
(for details see the original study~\cite{soltwisch2017reconstructing}).
In this study, we focus on testing the polygonal approach,
so the optimization procedure was not used.
Instead, we take the best-fit solution from~\cite{soltwisch2017reconstructing},
construct a polygon from it, and perform the forward computation using the polygonal approach.
In our simulations we do not consider the divergence of the beam. This can be done by convolving multiple solutions with the instrumental function of the incident beam. It has been shown that the far-field solution is sensitive to the horizontal divergence on the scale of $0.1^\circ$~\cite{Fernandez_Herrero2021}.
Therefore, it is important to consider the divergence of the incident beam for quantitative reconstruction of a structure. However, in our study we aim to validate the applicability of the polygonal approach for reconstruction where quantitative agreement is sufficient.
This allows us to verify the applicability of the polygonal approach to analyzing actual nanostructures.

The GISAXS patterns are measured at the BESSY-II synchrotron at the four-crystal monochromator beamline~\cite{krumrey2001high}.
The silicon grating sample with 150~nm pitch is prepared using electron-beam lithography.
Scattering patterns were measured with a fixed grazing incidence angle of
$\alpha_i = 0.86^\circ$ and at incident photon energy $\hbar\omega$ varying from 5500~eV to 5750~eV.
A single GISAXS pattern for $\hbar\omega = 5500$~eV is shown in Fig.~\ref{fig:pattern}(b).
Notice the bright diffraction peaks situated around the circular section.
The average structure of the grating's line defines the intensity distribution along these diffraction peaks.
In order to extract diffraction intensity distribution,
we integrate the pattern along the $q_y$ direction in a thin (seven pixels in width) stripe covering the cone section.
The resulting line cut with respect to $q_x$ is shown in Fig~\ref{fig:pattern}(a).
The final step in preparing the data is to integrate the cone section near the diffraction peaks and,
thus, to obtain a set of the total diffraction intensities.

Carrying out this procedure,
we extract the diffraction intensities from six GISAXS patterns measured for different $\hbar\omega$.
Then, we simulate the diffraction with the polygonal approach.
To do so, we construct a polygon by the best-fit model from.
Note that keeping the counterclockwise order of vertices is crucial for the numerical algorithm.
A polygon with 75 vertices is used for the simulations.
The computational domain in the reciprocal space is $81 \times 41$~nodes.
Amplitudes $R_a$ (from the block-vector $E_a$ in Eq.~\ref{eq:bdc}) of the Bloch wave in the ambient above the sample surface were calculated using the polygonal approach.
\begin{equation*}
	\mathcal{I}_n = \dfrac{k_{zn}}{k_{z0}}
	|R_n|^2 \exp(-\sigma^2 q_x^2). 
\end{equation*}
%
To compare the results with the experiment, we take into account the flux normalization.
The surface roughness was also taken into account, with the Deby-Waller damping factor.
The value $\sigma = 1.87$~nm of the best-fit roughness amplitude was again taken from~\cite{soltwisch2017reconstructing} and fixed.

\begin{figure*}[t]
	\center
	\includegraphics[width = 1\textwidth]{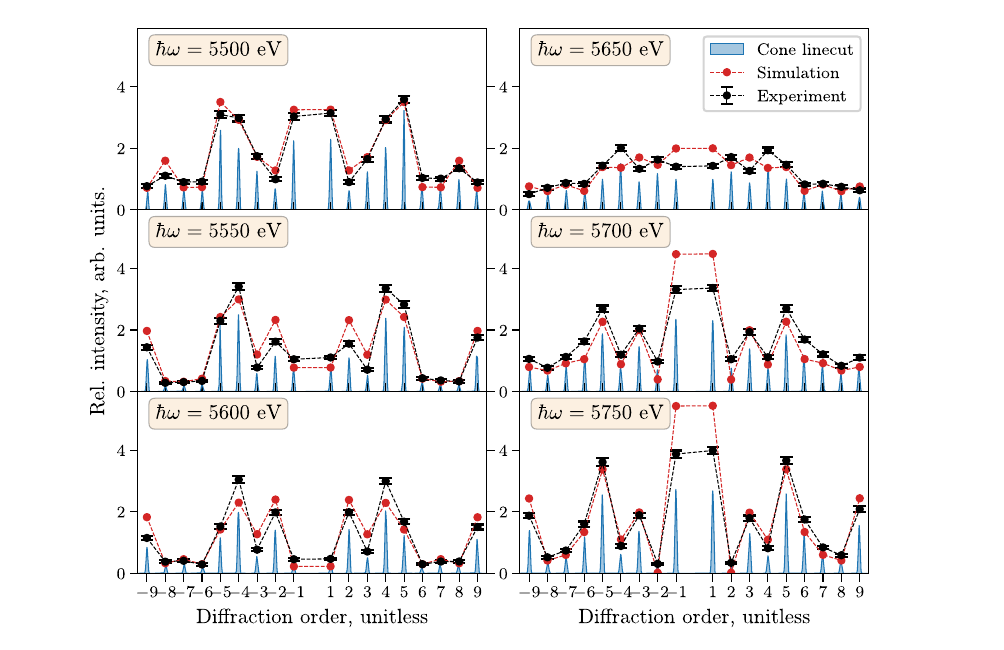}
	\caption{Comparison of diffraction intensity extracted from GISAXS patterns (black markers)
	with the polygonal approach simulation (red markers).
	The corresponding line cuts along the diffraction cone are shown as blue lines (not to scale).
	Six measurements with photon energies ranging from 5500 eV to 5750 eV are considered.}
	\label{fig:fit}
\end{figure*}

The comparison of the forward simulation with the experimental data is shown in Fig.~\ref{fig:fit}.
The experimental data are shown with black markers connected with the dashed line for the visual aid.
The error bars are calculated as a 3$\sigma$ of the shot-noise.
Any other kinds of errors are not accounted for.
The results of the simulation are in considerable agreement with the experimental data.
The line profile -- the upper part of a polygon used for the simulation,
is shown against the SEM image~\cite{soltwisch2017reconstructing} of the sample in Fig.~\ref{fig:tem_nf}(a).
They are in good visual agreement.
The near-field corresponding to a $\hbar\omega = 5500$~eV and $\alpha_i = 0.86^\circ$ incidence is in Fig.~\ref{fig:tem_nf}(b).

\section{Discussion and outlook}

With a new polygonal approach, we reproduced the results of FEM and RCWA on the test model computations.
Moreover, we simulated the X-ray diffraction from an actual sample,
achieving a reasonable agreement with the experiment.
Hence, the viability of using the new approach for reconstructing the parameters of nanostructures from the scattering patterns was demonstrated.
This opens a compelling opportunity for applications in synchrotron nanometrology, namely the free-form reconstruction.

In this context, the free-form approach implies the possibility of reconstructing the nanostructure's geometry without prior knowledge of its shape, nor even the knowledge to which class of shapes this structure's geometry belongs.
Nowadays, the free-form approach is realized for one-dimensional structures~\cite{zameshin2016reconstruction} only,
namely the thin films and the multilayers.
The 1D free-form approach is used to reconstruct the average density profile of a nanostructure with respect to the depth.
Without a free-form approach, one would have to define the class of function that describes the density gradient in depth.
Then, if a wrong function was chosen,
the reconstruction would be unsuccessful.
The free-form approach does not rely on any analytic expression of the density profile and can handle any profile shape with a predefined resolution.

Naturally, extending the free-form approach to higher-dimensional structures is not trivial.
When analyzing nanoscale gratings, one is restrained to assume a geometry model of the grating's line.
Assume for instance, one considers a simplistic square model of a line,
whereas the actual line has inclined walls.
Then, one modifies their model to include the sidewall angle.
However, assume, the actual structure is even more complex and has corner roundings.
Then, one has to modify the model once again, and so on.
Indeed, such a fitting process can not be done seamlessly.
At each step of increasing the reconstruction complexity, one has to devise a new model.
With our polygonal approach, this process can be carried out much easier.
First, one considers a square polygon and allows the walls' position to be free parameters for the fit.
If the square model is not adequate for the actual structure, one allows the position of each vertex to be fitted.
Then, the tilted sidewalls automatically appear in the model.
If the structure is even more complex, one should only increase the number of vertices.
The only restrictions for the model are that the vertices must be ordered counterclockwise and the polygon must not cross itself.

An obstacle to realizing a free-form analysis based on the polygonal approach is the computation speed.
The conventional RCWA approach is based on a one-dimensional Fourier transform of the scattering potential,
while the new approach assumes a two-dimensional structure.
this leads to the eigenproblem of a larger size (cf. Eq.~\ref{eq:qep}) and necessity to solve a QEP.
The computational complexity problem can be circumvented by developing an algorithm to find fast approximate solutions to the eigenvalue problems,
which takes into account the matrix structure of the QEP.
Indeed, the information entropy of the characteristic matrix in our case is reduced compared to an arbitrary matrix of the same size,
as the arising matrix $C_{[\chi]}$ is Toeplitz
(has entries constant along diagonals).
It hints at the possibility of devising a numerically effective algorithm specific to our type of QEP.
We can restate that in terms of physical intuition.
The farther the wave modes are separated in the reciprocal space (the difference between diffraction order numbers), the less these modes interfere.
Thus, there is no need to take into account the interference between each mode.
Hence, the approximate solution should exist.

If there is an approximate solution, parity between decay and gain modes may no longer be guaranteed, as in the cases described in~\cite{medvedev2006anomalous, nikolaev2018specular}.
In this case, the asymmetric BDC solution Eq.~\ref{eq:bdc_new} could become useful.
We focus our future research on finding a numerically efficient solution and realizing a free-form approach for reconstructing the grating's structure.

\section{Conclusion}
	We have described a Fourier modal approach to the simulation of scattering in lamellar gratings. It is based on a combination of ideas from rigorous coupled wave analysis and dynamical diffraction theory. This allowed us to consider the grating as it is, without approximating it with layered models and without constraining the many beam interactions. We also implemented polygons to model the grating profile. This in turn allows us to easily calculate scattering on gratings of arbitrary complexity. This is a substantial improvement over a conventional approach in which the profile geometry model is fixed and any new shape requires a new model to be devised. We have compared the new approach with rigorous coupled wave analysis and the finite element method. All considered approaches produce solutions of similar order of accuracy with relative $L_2$ norms less than 2\% compared to the finite element solution. Thus, we conclude that the new approach is suitable for quantitative simulations. Furthermore, we have compared the simulations performed by the new approach with the synchrotron scattering experiments. The simulations are in qualitative agreement with the experiment. The new approach can be employed as a nanometrology tool for future microelectronics and in other fields of science and technology. This will require further research to improve the computational speed. However, the present study takes the first steps in this direction. Namely, the generalized solution of the boundary condition problem has been found, which may allow the implementation of faster algorithms in the future.

\section*{Acknowledgments}
    The authors are grateful to Vladimir Kaganer for the sharing of his expertise and for fruitful discussions.
        
\bibliography{bibliography}{}
\bibliographystyle{ieeetr}

\end{document}